# Blue luminescence of Au nanoclusters embedded in silica matrix

S. Dhara,[a)] Sharat Chandra, P. Magudapathy, S. Kalavathi, B. K. Panigrahi, K.G.M. Nair and

V.S. Sastry

*Materials Science Division, Indira Gandhi Centre for Atomic Research, Kalpakkam-603 102, India*

C. W. Hsu, C.T. Wu and K.H. Chen

*Institute of Atomic and Molecular Sciences, Academia Sinica, Taipei 106, Taiwan*

L.C. Chen

*Center for Condensed Matter Sciences, National Taiwan University, Taipei 106, Taiwan*

## ABSTRACT

Photoluminescence study using the 325 nm He-Cd excitation is reported for the Au nanoclusters embedded in $SiO_2$ matrix. Au clusters are grown by ion beam mixing with 100 KeV $Ar^+$ irradiation on Au [40 nm]/$SiO_2$ at various fluences and subsequent annealing at high temperature. The blue bands above ~3 eV match closely with reported values for colloidal Au nanoclusters and supported Au nanoislands. Radiative recombination of *sp* electrons above Fermi level to occupied *d*-band holes are assigned for observed luminescence peaks. Peaks at 3.1 eV and 3.4 eV are correlated to energy gaps at the *X*- and *L*-symmetry points, respectively, with possible involvement of relaxation mechanism. The blue shift of peak positions at 3.4 eV with decreasing cluster size is reported to be due to the compressive strain in small clusters. A first principle calculation based on density functional theory using the full potential linear augmented plane wave plus local orbitals (FP-LAPW+LO) formalism with generalized gradient approximation (GGA) for the exchange correlation energy is used to estimate the band gaps at the *X*- and *L*-symmetry points by calculating the band structures and joint density of states (JDOS) for different strain values in order to explain the blueshift of ~0.1 eV with decreasing cluster size around *L*-symmetry point.

---

For any Correspondence Email : dhara@igcar.ernet.in

**INTRODUCTION**

Optical properties of noble metal nanoclusters embedded in the dielectric matrices are of current interest due to their potential applications in nonlinear optics.[1] Surface plasmon resonance (SPR) for embedded noble metal (e.g., Au, Ag and Cu) nanoclusters in the visible region has attracted special attention.[2] Photoluminescence (PL) of Au nanoclusters is at centre of the discussion regarding its origin and optoelectronic device applications. Broad red luminescence around ~2 eV is reported for Au nanoislands (≤10 nm) supported on rough $SiO_2$ surface.[3] Radiative decay of collective plasmon states of the nanoislands is thought to be responsible for the observed PL. Blue luminescence from both the supported Au nanoislands (≥ 10 nm)[4] and Au colloidal suspension (~5 nm)[5] have also been reported. The origin of PL is primarily assigned to the radiative recombination of Fermi level electrons and $d$-band holes.[4,5] Moreover, recent studies on lifetime of hot electrons (electrons above Fermi level) in noble metals, namely, Au and Cu have shown interesting results with increased decay time considering electron-electron ($e$-$e$) scattering involved in the relaxation process.[6,7] With several advantages of embedded nanoclusters, e.g., its chemical stability and capability of use in practical nonlinear device applications, PL study of these structures is of high importance.

In this study, we report the observation of PL from embedded Au nanoclusters formed by ion beam mixing method using 100 keV $Ar^+$ in $Au/SiO_2$ at various fluences and subsequent high temperature annealing. Assuming strain as a parameter, a first principle calculation was carried out using the WIEN2k package.[8] This is a density functional theory (DFT) code, which uses the full potential linear augmented plane wave plus local orbitals (FP-LAPW+LO) formalism with generalized gradient approximation for the exchange correlation energy (GGA-96)[9] for calculating the crystal properties. The linear optical properties like joint density of states (JDOS) and dielectric constant have been calculated using the random phase approximation (RPA) taking into account inter- as well as intra-band contributions.[10] These calculations have been used to understand the changes in the peak positions of blue bands with strain in the nanoclusters. Optical absorption studies using UV-VIS spectroscopy were performed to determine the size of Au nanoclusters. Volume strain in the small clusters was calculated using the percentage change in the lattice parameters derived from the glancing incidence x-ray diffraction studies on the samples. Prominent blue luminescence for the Au nanoclusters grown at various fluences was recorded using a 325 nm He-Cd laser excitation.

**EXPERIMENTAL DETAILS**

Au clusters were grown by ion beam irradiation on Au [40 nm]/$SiO_2$. Thin film of Au was coated on silica substrates by thermal evaporation technique at 0.2 nm/sec deposition rate with a base pressure of $1 \times 10^{-4}$ Pa. The films were irradiated using 100 keV $Ar^+$ at $5 \times 10^{-5}$ Pa with a beam current of $1 \times 10^{-2}$ A/$m^2$ to avoid sputtering effect in the fluence range of $1 \times 10^{20}$-$5 \times 10^{20}$ $m^{-2}$. Ion beam energy was chosen based on SRIM-2000 calculations to ensure that the maximum of the damage profile remained close to the Au/$SiO_2$ interface. The samples were annealed at a temperature of 1173K in air for 1 hour for the growth of the clusters. Details of sample preparation are reported elsewhere.[11]

UV-VIS spectroscopic measurements (Shimadzu PC 3101) were performed in the energy range of 1.5 - 3.5 eV. Transmission electron microscopy (TEM; Philips 200) study was used for the direct observation of typical nanoclusters. Glancing incidence X-ray diffraction (GIXRD) measurements were carried out using a STOE diffractometer for finding the lattice parameter of the nanoclusters as well as to corroborate the particle sizes. PL measurements were performed using He-Cd laser tuned to 325 nm with an output power of ~10 mW at room temperature. The emission signal was collected in back scattered geometry by a SPEX 0.85-m double spectrometer and detected by a lock-in-amplifier.

**RESULTS AND DISCUSSION**

Post-irradiation annealing at 1173K showed strong SPR peaks around 2.5 eV (Fig. 1), which are characteristic of Au nanoclusters embedded in $SiO_2$. Sharp SPR peaks were observed in the UV-Vis spectra for the annealed samples with peak positions shifting towards higher energy with decreasing fluence. Assuming the free particle behavior for conduction electrons in Au nanoclusters, the average cluster diameter of gold was calculated with $<D> = hV_F /2\pi\Delta E_{1/2}$ where $V_F$ (=1.395x10$^6$ m sec$^{-1}$) is the Fermi velocity of electrons in bulk gold, $h$ is the Planck's constant and $\Delta E_{1/2}$ is the full-width at half-maximum of the absorption band. Cluster sizes (<10 nm) are observed to increase with increasing fluence (inscribed in Fig. 1). Formation of spherical clusters with average diameter $<D> = 7.7 \pm 0.2$ nm for 1173K annealed samples are shown (Fig. 2) in the planar TEM micrograph, for the typical sample grown at a fluence of 5x10$^{20}$ m$^{-2}$. The particle size matches well with that calculated from the FWHM of UV-VIS absorption spectra of the corresponding samples. A reasonably narrow size distribution for the clusters is also observed in these samples (inset Fig. 2). Nucleation of gold atoms occurs with the low concentration fluctuations, and these nuclei grow directly from the supersaturation (due to poor solubility of Au in SiO$_2$) leading to agglomeration of bigger clusters with increasing fluence.[12] Growth of the cluster size with increasing annealing temperature may be due the fact that nuclei, which have reached a critical size, grow at the expense of smaller nuclei, known as the coarsening or ripening stage.[12] A blueshift of SPR peak with decreasing cluster size is observed (inset in Fig. 1) in our study. The physical idea underlying the blueshift trend with decreasing cluster size observed in noble metal clusters is based on the assumption that the screening effects are reduced over a surface layer inside the metallic particle due to the localized character of the core-electron wavefunctions. Close to the surface, the valence electrons are then incompletely embedded inside the ionic-core background and core-electrons contribute more to the metal dielectric function, giving rise to the blueshift with decreasing cluster size (inset in Fig. 1).[11,13]

PL spectra for the post-irradiated annealed samples show (Fig 3) two major peaks above 3 eV. Broad blue bands at ~3.1 eV and ~3.4 eV match closely with those reported for ~5 nm Au colloidal solution.[5] A prominent peak at 3.1 eV is also reported for supported Au nanoislands with relatively larger size (≥10 nm).[4] A small blueshift (~0.1 eV) along with increase in the intensity is observed for peak position at 3.4 eV with decreasing fluence, i.e., decreasing nanoparticle size. The peak at 3.1 eV is observed to dominate for the sample irradiated at a high fluence of $5 \times 10^{20}$ m$^{-2}$ and no appreciable change in the peak position is observed. Intensity of a peak at ~1.74 eV is observed to decrease with increasing fluence.

We resort to the mechanism of excitation of electron-hole (*e-h*) pair, subsequent relaxation and finally radiative recombination of electrons above Fermi level to occupied *d*-band holes for the observed PL peaks above 3 eV.[4,5] Calculated single photon excitation spectra predicted the peak at 3.1 eV to be a mixture of $L_6 \circledR L_4$ and $X_6 \circledR X_3$ transitions (states 6 and 3, 4 correspond to *sp* conduction electrons above Fermi level and *d*-band holes, respectively). The peak at 3.4 eV was also predicted to arise from $L_6 \circledR L_4$ transition alone, but was not observed experimentally for large (≥10 nm) Au cluster size.[4] A small blueshift of the 3.4 eV peak with decreasing cluster size (Fig. 3) may be related to the strain in the lattice of small clusters. When the spherical gold nanoparticles are formed in the SiO$_2$ matrix, the matrix exerts an isotropic pressure on the gold nanoparticle, thus reducing the average lattice constant in gold by a small percentage and giving rise to the compressive strain observed in the gold lattice. Typical Glancing incidence x-ray diffraction (GIXRD) study (Fig. 4) was performed for determining the volume strain in the post-annealed samples. An average compressive strain up to ~1.1% was calculated for the cluster size of ~ 7.3 nm grown at a fluence of $3 \times 10^{20}$ m$^{-2}$. We could not get meaningful data from the GIXRD analysis of the sample with cluster size smaller than ~7.3 nm.

We have attempted to explain the observed optical properties on the basis of the calculations carried out on the bulk gold. In our opinion this is justified for the following reasons. The sizes of the nanoparticles that are under study are much larger than the Bohr radius for the exciton (~0.5 nm) and contain a very large number of gold atoms. This implies that we are far away from the confinement regime and are working in the mesoscopic range where the properties are between atom-like and bulk-like. Thus a majority of the observed optical properties can be explained by invoking the bulk properties. The only difference would be that the calculated transitions would be suitably modified because of the breakdown of the infinite crystal approximation. Assuming strain to be highest in smallest size cluster, we have carried out fully relativistic calculations of the band structures of Au with strain as a parameter using DFT with GGA formalism. First, the unit cell configuration with minimum energy was found by optimizing the Au structure. This corresponded to a energy optimized lattice constant of 7.66 Bohr (4.0533 Å). Then the properties of the Au structure under compressive strain were calculated by varying the volume of the optimized unit cell by a small percentage corresponding to 0.0, 0.2, 0.5, 0.8, 1.0, 1.2, 1.5 and 1.8% change in volume. Self-consistency cycles were performed with 72 k-points in the irreducible Brillouin zone (IBZ). The Brillouin zone integrations for obtaining JDOS and the dielectric constants were carried out with 1240 k-points in the IBZ, or 50,000 k-points in the full Brillouin zone for all the strain configurations. Convergences were carried out both with respect to the energy and charge and the final convergences obtained were $6 \times 10^{-6}$ for the energy and $3 \times 10^7$ for the charge. The *5s*, *5p* and *4f* states in the valence region were treated by local orbitals. Spin-orbit coupling becomes important in the relativistic case and it was taken into account during the self-consistency cycles. JDOS and dielectric constants were calculated taking into account both the interband and intraband contributions, but it was found that the intraband effects were important only below 2 eV. Hence all the results have been presented for the interband contributions only. The calculated band structure with zero and -1.8% strains is shown in Fig. 5. The bands around the Fermi energy have been numbered by counting the bands from the deepest energy valence band. The valence bands including band 8 are dominated by the *d*-character. Band 9 has *pd*-character in *W-L* direction; *d*-character in the *L-G-X* direction and *p*-character in *X-W* direction. Band 10 has *sd*-character in *W-L* direction; *f*-character around the *G*-point and *s*-character in *X-W* direction.[10] We observe that the band number 9 crosses the Fermi energy and that the transitions take place along the *X*- and *L*-symmetry points between bands 9 and 10 at the *L*-symmetry point and between band 8 and 9 at the *X*-symmetry point. In all the cases, the initial states are mainly *d*-bands, while the final states range from the Fermi level up to very high energies.[10] It can also be seen that there is not much difference in the band structure due to the presence of strain. The calculated band gaps correspond to the transitions along the *X*- and *L*- symmetry points. Here it is necessary to note that no ground state calculations including exchange interactions can properly account for all the possible relaxation mechanisms involved in the transition process.[4,5]

We have calculated the JDOS for understanding the observed optical properties which when multiplied by the transition matrix elements gives the imaginary part of the dielectric constant (*Im e*). The imaginary part of the dielectric constant is already well estimated by the JDOS and the matrix element effect is relatively small.[10] Since the dielectric function suppresses higher energy transitions due to the $1/\omega^2$ behavior, we have plotted in Fig.6 the JDOS/$E^2$ versus energy (E) in eV in order to present the variations in JDOS clearly. Inset of the Fig.6 shows the variation in JDOS/$E^2$ in 0-8 eV energy range for the optimized gold structure. The calculated *Im e* is plotted in Fig.7 for all the strain values. The inset in Fig.7 shows the real (*Re e*) and *Im e* for the optimized gold structure in 0-8 eV energy range. We have calculated the position of SPR peak in the framework of Mie scattering theory[13] using the calculated energy dependent dielectric constant data. A SPR peak is observed for Au nanoclusters in silica matrix around ~2.5 eV depending on the cluster size. Hence the peak around ~2.6 eV (Fig. 7) can be attributed to the plasmon resonance, signature of which is observed in Fig. 1. Absorption around ~2.6 eV is dominated by non-radiative recombination processes and does not contribute to the PL process. The sharp feature at ~5 eV in JDOS is not observed in the *Im e* plot due to cancellation by the selection rules. Here it is necessary to note that the selection rules are modified when we are studying the nanoparticle spectra due to the breakdown of the infinite crystal approximation. The two transitions at ~3.1 and ~3.4 eV are marked by arrows in Fig.7 and correspond to broad shoulders in the *Im e* spectra. The transitions at ~3.4 eV show a blueshift as the strain increases or the particle size decreases. This corresponds to a blue shift of ~0.1 eV between the compressive strain values of 1.8% (assumed for ~5.7 nm diameter) and 0.6% (observed for ~7.7 nm diameter). The broad shoulder also becomes increasingly well defined with decreasing nanoparticle size, which can lead to an increase in the magnitude of the transition. The blue shift with increasing strain (decreasing cluster size) supports our observation for the PL peak at 3.4 eV (Fig. 3). Fig.7 also shows that the peak position at ~3.1 eV does not change appreciably with decreasing nanoparticle size or increasing compressive strain: a behavior similar to that observed in the PL peak at 3.1 eV (Fig.3). The changes in the magnitudes of the PL intensities in the experimental observation (Fig.3) could not be adequately explained by our calculations. These changes in intensities of PL peaks with cluster size may not be correlated only to the *Im e* magnitude but relaxation mechanisms are also involved in the process and these are not considered in our band structure calculations.[9] In other words, at present, we have not understood the whole transition process, in particular the change in intensities with the cluster size.

There can be many contributions for the peak observed at ~1.74 eV. It can have contributions from the first interband transition present at ~1.75 eV in the dielectric spectra of gold corresponding to the absorption edge.[10] The absorption edge has not been observed to shift with increasing strain in our calculations. Also the nonbridging oxygen (NBO) defects which are present in mesoporous $SiO_2$ give rise to transitions in this energy range.[14] The NBO defects arise due to breaking of bonds in the energetic irradiation process. Increasing amount of metallic Au with increasing fluence may contribute to the reduction of NBO defects in $SiO_2$ and thus reduces the intensity of the peak at 1.74 eV. As a matter of fact, oxygen vacancy related defects are reported to be the active nucleation sites for Au clusters in the dielectric matrices.[15]

**CONCLUSIONS**

Blue luminescence of Au nanoclusters (<10 nm) embedded in $SiO_2$ matrix is reported at 3.1 eV and 3.4 eV. Radiative recombination followed by a relaxation process of *sp* conduction electrons above Fermi level and *d*-band holes is assigned for the observed luminescence peaks. Increasing compressive strain with decreasing cluster size is correlated with the blueshift of peak at 3.4 eV with decreasing cluster sizes. Understanding of the whole transition process, particularly the change in intensities with cluster size, is not complete in the present study.

**ACKNOWLEDGEMENTS**

The authors would like to thank M. C. Valsakumar for providing valuable inputs in band structure calculations and careful reading of the manuscript. We also thank M. Premila for carrying out the UV-VIS spectroscopic measurements.

**REFERENCES**

[1] H.B. Liao, R.F. Xiao, J.S. Fu, P. Yu, G.K.L. Wong, P. Sheng, Appl. Phys. Lett. **70**, 1 (1997).

[2] U. Kreibig, M. Vollmer, *Optical Properties of Metal Clusters* (Springer series, Berlin, 1995).

[3] L. Khriachtchev, L. Heikkila, and T. Kuusela, Appl. Phys. Lett. **78**, 1994 (2001).

[4] G.T. Boyd, Z.H. Yu, and Y.R. Shen, Phys. Rev. B **33**, 7923 (1986).

[5] J.P. Wilcoxon, J.E. Martin, F. Parsapour, B. Wiedenman and D.F. Kelley, J. Chem. Phys. **108**, 9137 (1998).

[6] I. Campillo, J.M. Pitarke, A. Rubio, and P.M. Echenique, Phys. Rev. B **62**, 1500 (2000); J. Cao, Y. Gao, H.E. Elsayed-Ali, R.J.D. Miller and D.A. Mantell Phys. Rev. B **58**, 10948 (1998).

[7] V.P. Zhukov, F. Aryasetiawan, E.V. Chulkov, I.G. de Gurtubay, and P.M. Echenique, Phys. Rev. B **64**, 195122 (2001).

[8] P. Blaha, K. Schwarz, G.K.H. Madsen, D. Kvasnicka and J. Luitz, *WIEN2k, An Augmented Plane Wave + Local Orbitals Program for Calculating Crystal Properties* (Karlheinz Schwarz, Tecn. Universität Wien, Austria), 2001. ISBN 3-9501031-1-2

[9] J.P. Perdew, K. Burke, M. Ernzerhof, Phys. Rev. Lett. **77**, 3865 (1996)

[10] C. Ambrosch-Draxl and J.O. Sofo, http://arXiv.org/abs/cond-mat/0402523

[11] S. Dhara, R. Kesavamoorthy, P. Magudapathy, M. Premila, B.K. Panigrahi, K.G.M. Nair, C.T. Wu, K.H. Chen, and L.C. Chen, Chem. Phys. Lett. **370**, 254 (2003).

[12] P. Sigmund, and A. Gras-Marti, Nucl. Instr. and Meth. in Phys. Res. B **182/183**, 25 (1981).

[13] J. Lermé, B. Palpant, B. Prével, E. Cottancin, M. Pellarin, M. Treilleux, J.L. Vialle, A. Perez, and M. Broyer, Eur. Phys. J. D **4**, 95 (1998).

[14]A.S. Zyubin, Yu.D. Glinka, A.M. Mebel, S. H. Lin, L.P. Hwang, and Y.T. Chen, J. Chem. Phys. **116**, 281 (2002).

[15]E. Wahlström, N. Lopez, R. Schaub, P. Thostrup, A. Rønnau, C. Africh, E. Lægsgaard, J. K. Phys. Rev. Lett. **90**, 26101 (2003); J. Xu, A.P. Mills Jr., A. Ueda, D.O. Henderson, R. Suzuki, and S. Ishibashi, Phys. Rev. Lett. **83**, 4586 (1999).

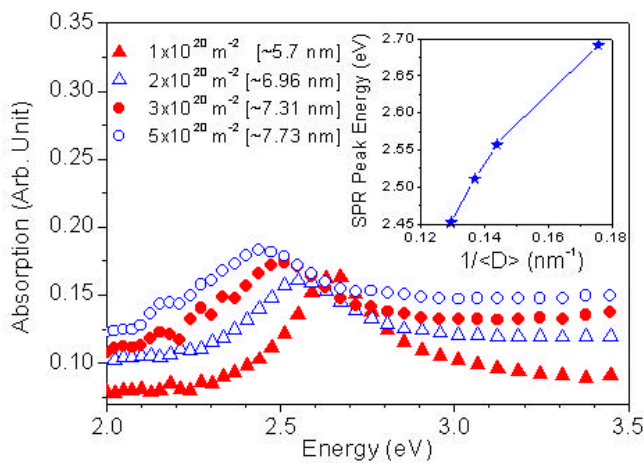

Fig. 1. SPR peaks for the annealed samples grown at various fluences. The spectra are vertically shifted for clarity. Inset shows the blueshift with decreasing fluence.

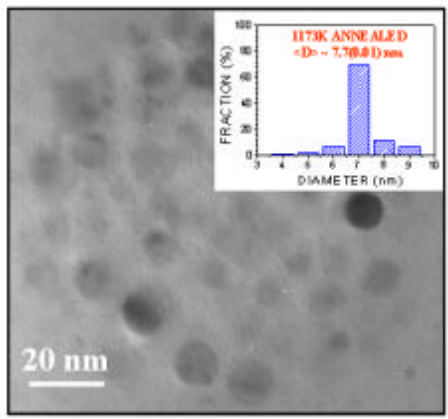

Fig. 2. Transmission electron micrograph of typical Au clusters grown at a fluence of $5 \times 10^{20}$ m$^{-2}$ and annealed at 1173K. Inset shows the size distribution of clusters.

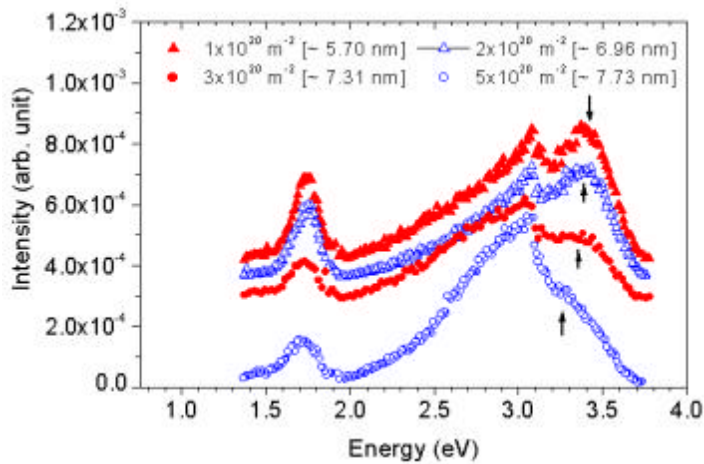

Fig. 3. PL spectra for the annealed samples grown at various fluences obtained with the 325 nm (~3.8 eV) He-Cd excitation. Cluster sizes corresponding to the various fluences are also given. A small blueshift (~0.1 eV) is observed for the peak at 3.4 eV with decreasing growth fluence, i.e., decreasing cluster size. Peaks are shifted vertically for clarity. Shifting peak (around 3.4 eV) positions with fluences are shown by arrows.

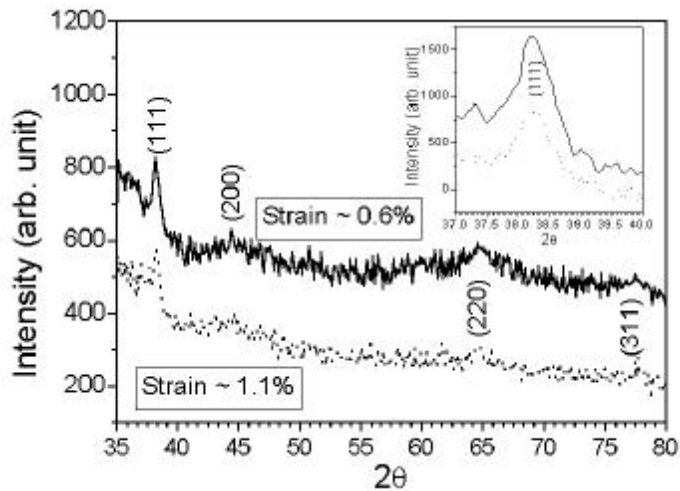

Fig. 4. GIXRD plots of the post-annealed samples grown at fluences of $3 \times 10^{20}$ (dotted line) and $5 \times 10^{20}$ m$^{-2}$ (continuous line). Peak positions are shifted towards higher angle than that for the bulk value, indicating shortening of lattice parameter and thereby the presence of compressive volume strain in the small clusters, which is also indicated in the figure. Inset shows peaks assigned to (111) plane with larger accumulation time for corresponding samples.

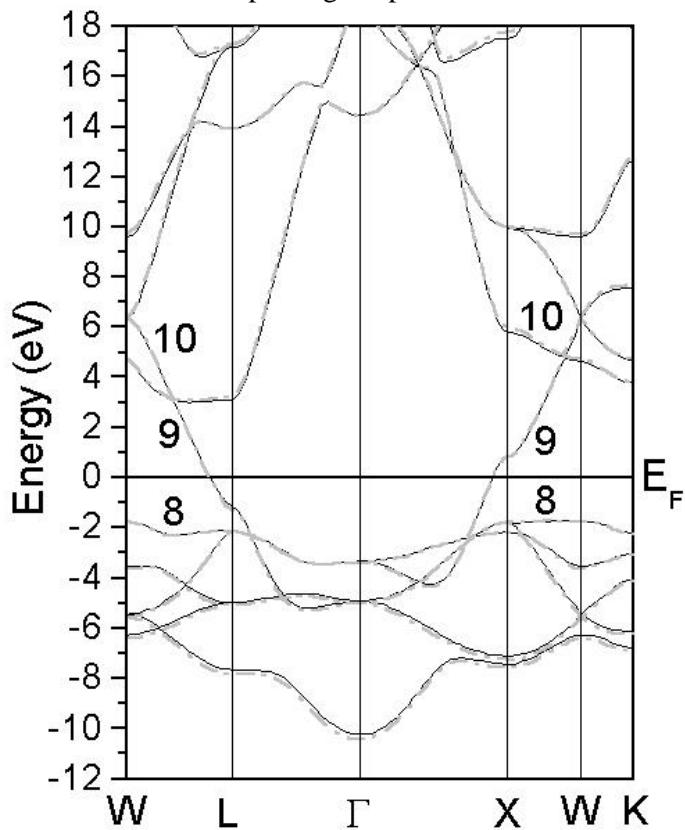

Fig. 5. Band structure of Au calculated along certain symmetry directions. The band structures for the 0.0% (solid black line) and -1.8% (dashed grey line) strains are shown plotted in the same figure. The bands are numbered as counted from the deepest valence band.

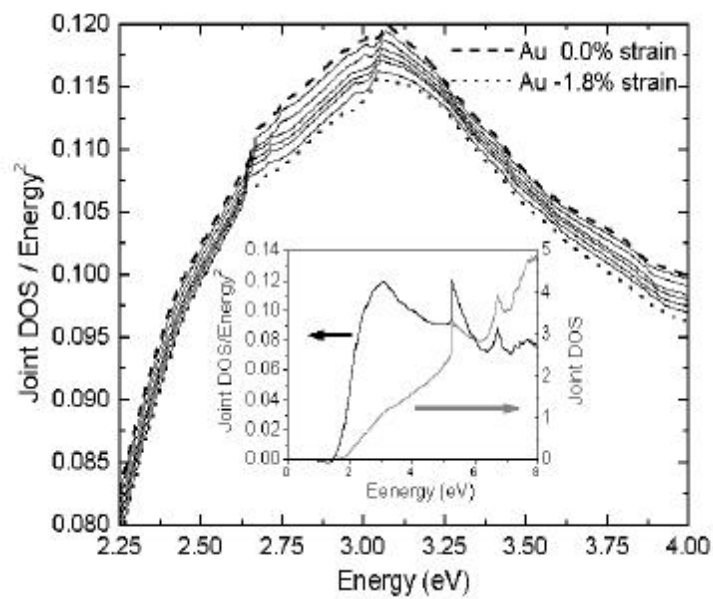

Fig. 6. Joint DOS of gold for various compressive strain values obtained using DFT with GGA. Inset shows the Joint DOS plotted for the unstrained optimized gold in 0-8 eV energy range.

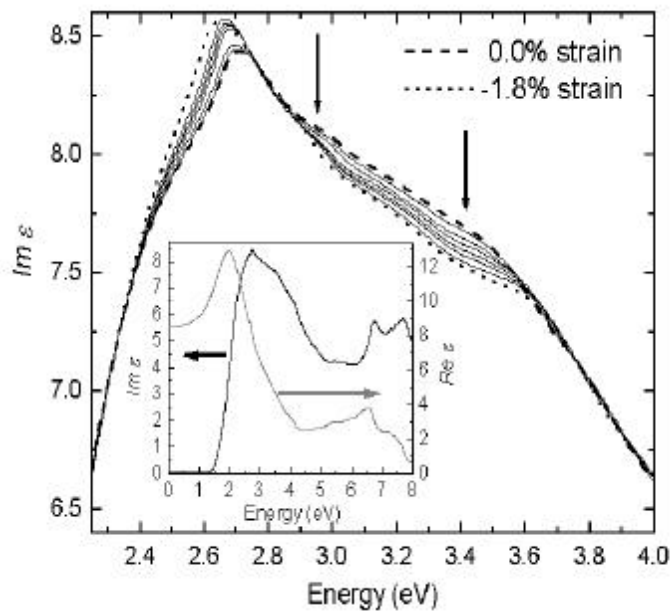

Fig. 7. Imaginary part of the calculated interband dielectric constant of Au plotted for various strain values. The inset shows the real (solid grey line) and imaginary (solid black line) parts of the interband dielectric constant for the unstrained optimized gold in 0-8 eV energy range.

13